\documentclass[letterpaper, 10 pt, conference]{ieeeconf}  

\IEEEoverridecommandlockouts                              
                                                          
\usepackage{cite}
\usepackage{amsmath,amssymb,amsfonts}
\usepackage{algorithmic}
\usepackage{graphicx}
\usepackage{textcomp}
\usepackage{xcolor}
\usepackage{soul}
\usepackage{tikz}
\usepackage{hyperref}
\usepackage{xspace}
\usepackage[font=small]{caption}
\usepackage{subcaption}
\def\BibTeX{{\rm B\kern-.05em{\sc i\kern-.025em b}\kern-.08em
    T\kern-.1667em\lower.7ex\hbox{E}\kern-.125emX}}

\definecolor{color1}{RGB}{166, 206, 227}
\definecolor{color2}{RGB}{ 31, 120, 180}
\definecolor{color3}{RGB}{178, 223, 138}
\definecolor{color4}{RGB}{ 51, 160,  44}
\definecolor{color5}{RGB}{251, 154, 153}
\definecolor{color6}{RGB}{227,  26,  28}
\definecolor{color7}{RGB}{253, 191, 111}
\definecolor{color8}{RGB}{255, 127,   0}

\begin{document}

\title{\LARGE \bf Non-local Evasive Overtaking of Downstream Incidents in Distributed Behavior Planning of Connected Vehicles}

\newcommand{\methodName}[0]{NEO\xspace}
\newcommand{\benchmarkName}[0]{Altruistic MOBIL\xspace}

\author{Abdul Rahman Kreidieh$^{1,2}$, Yashar Farid$^{1}$, and Kentaro Oguchi$^{1}$
\thanks{$^{1}$Toyota InfoTech Labs}%
\thanks{$^{2}$University of California, Berkeley}%
}

\maketitle

\begin{abstract}
The prevalence of high-speed vehicle-to-everything (V2X) communication will likely significantly influence the future of vehicle autonomy. In several autonomous driving applications, however, the role such systems will play is seldom understood. In this paper, we explore the role of communication signals in enhancing the performance of lane change assistance systems in situations where downstream bottlenecks restrict the mobility of a few lanes. Building off of prior work on modeling lane change incentives, we design a controller that 1) encourages automated vehicles to subvert lanes in which distant downstream delays are likely to occur, while also 2) ignoring greedy local incentives when such delays are needed to maintain a specific route. Numerical results on different traffic conditions and penetration rates suggest that the model successfully subverts a significant portion of delays brought about by downstream bottlenecks, both globally and from the perspective of the controlled vehicles. 
\end{abstract}


\section{Introduction}

The adoption of vehicle autonomy and wireless communication is expected to redefine the nature of future transportation systems. In autonomous intersection scenarios, for instance, intelligent polling systems~\cite{miculescu2014polling, miculescu2019polling} and 
broadcasted
safety messages~\cite{kenney2011dedicated} are expected to improve the efficiency and safety of such systems, respectively. 
Similarly, in freeway
merging scenarios, coordinated 
acceleration 
behaviors by \emph{connected and automated vehicles} (CAVs) are expected to increase the throughput of such networks and reduce traffic instabilities~\cite{zhou2016impact, kreidieh2018dissipating}. 
As such, an 
in-depth 
understanding of the role of 
information sharing 
in all CAV applications is necessary to ensure that such systems thrive.


In the context of automated lane changing, the role of communication is less understood. This limits the efficacy of existing models, which attempt to balance personal and societal objectives using locally observable data, e.g. the state of neighboring vehicles. 
This limitation is particularly evident in congested settings, in which more egoistic lane change models often contribute to degradation in throughput (an issue previously noted for similar longitudinal controllers~\cite{gunter2020commercially}), while more altruistic models produce imbalances in performance between humans and CAVs (we illustrate this in Section~\ref{sec:results-disc}). 
Knowledge of the source of congestion, we hypothesize, may help alleviate these concerns.

\begin{figure}
\centering
\begin{subfigure}[b]{.49\textwidth}
	\includegraphics[width=\textwidth]{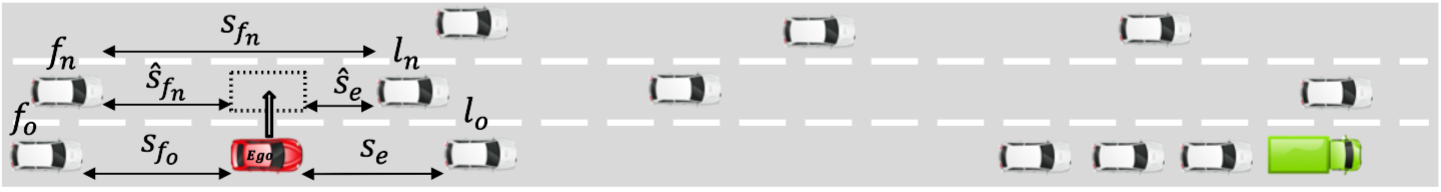}
	\caption{Lane change without detection\\[6pt]}
	\label{fig:discretionary_a}
\end{subfigure}
\begin{subfigure}[b]{.49\textwidth}
	\includegraphics[width=\textwidth]{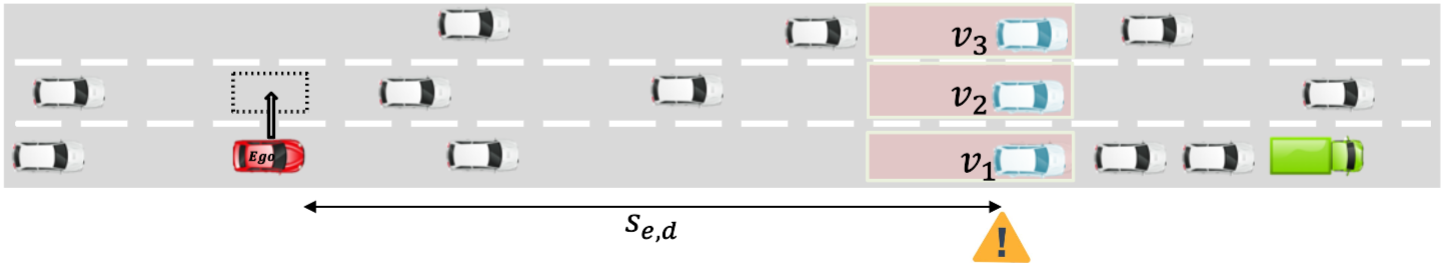}
    \caption{Discretionary lane change with detection\\[6pt]}
	\label{fig:discretionary_b}
\end{subfigure}
\begin{subfigure}[b]{.49\textwidth}
	\includegraphics[width=\textwidth]{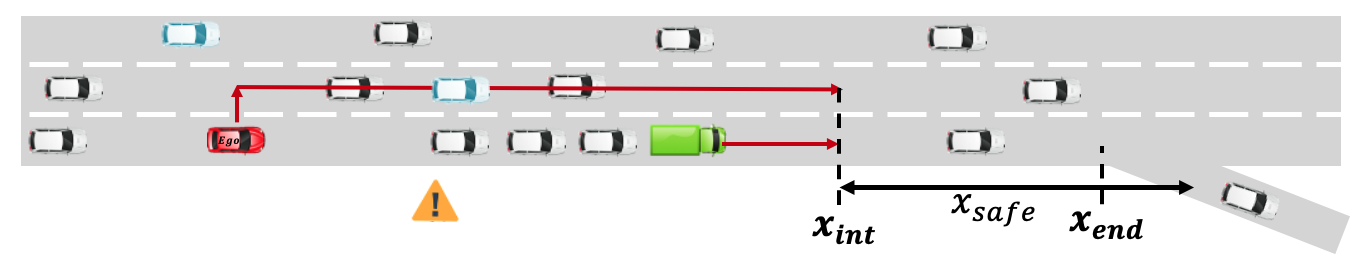}
    \caption{Mandatory lane change with detection}
	\label{fig:mandatory}
\end{subfigure}
\caption{An illustration of the proposed model. \textbf{a)} Classic models compute incentives based on local metrics, without accounting for communicable non-local features. \textbf{b)} We design an incentive-based model that accounts for speed reductions from distant downstream incidents. \textbf{c)} For mandatory lane change events, we further augment the incentives to overtake slow-moving vehicles in a desired lane.}
\vspace{-.5em}
\label{fig:intro}
\end{figure}

In this study, we aim to determine whether knowledge of the source of propagating disturbances can be exploited to improve the lane change behavior of CAVs. In particular, we explore the benefit of such systems in highway networks whereby throughput is restricted by the presence of a slow-moving, incident vehicle (Figure~\ref{fig:intro}). Assuming the incident is visible from a V2X sense, we augment the work of~\cite{kesting2007general} (see Section~\ref{sec:mobil}) by introducing lane change incentives that inform the CAV to avoid, and if possible, overtake existing downstream congestion. Our model for Non-local Evasive Overtaking of downstream incidents, or \methodName, is demonstrated to reduce the failure rate of CAVs attempting to reach certain routes in mandatory settings while improving the performance and reducing the disparity between humans and CAVs brought about by altruistic models in discretionary settings. Numerical results from various simulated errors also suggest that \methodName is robust to limitations in sensing and sharing.

The key contributions of the paper are:

\begin{itemize}
    \item We design an incentive-based model for lane changing that accounts for non-local communicable state information. Our model attempts to avoid lanes with distant downstream incidents and overtake or track such incidents when required to maintain a specific route.
    \item We compare the controller's performance to that of similar models that solely sense local data, demonstrating the model's efficacy in incident avoidance.
\end{itemize}

The remainder of this paper is organized as follows. Section~\ref{sec:related-work} provides an overview of existing lane change models and their relation to ours. Section~\ref{sec:method} introduces the proposed method for exploiting non-local speed-gain incentives to lane change decision-making. Section~\ref{sec:results} presents the findings and results of computational experiments over a variety of network configurations and traffic states and compares the performance of the proposed and existing models. Finally, Section~\ref{sec:conclusion} provides concluding remarks.  

\section{Related Works} \label{sec:related-work}

\subsection{Lane change models} \label{sec:lane-change-models}

Lane change models denote the decision-making behavior through which drivers choose to switch lanes. These models often distinguish lane changes as one of two types: 1) \emph{discretionary}, in which drivers change to a lane perceived to offer better traffic conditions, most notably faster driving speeds, and 2) \emph{mandatory}, in which drivers change lanes to follow a specified path, for example in the direction of an off-ramp. In both settings, particular focus is placed on estimating a given target lane's safety and utility (particularly in terms of speed gains and time-to-collision). For a more comprehensive review of lane change models, we refer the reader to~\cite{ahmed1999modeling, treiber2013traffic}.

The use of lane change models has evolved significantly over the years. In the past, such models served mainly as tools for simulating the behavior of lane changes in traffic settings~\cite{gipps1986model, kesting2007general, schakel2012integrated, erdmann2015sumo}, with a particular interest in their effects on traffic flow stability~\cite{ahn2007freeway, laval2006lane, zheng2011freeway}. With the advent of autonomous driving, however, much effort has been placed on re-purposing similar insights to augment the decision-making behaviors of automated vehicles (AVs). 
Most relevant to the present paper, models such as these have served to define the movement of AVs. These approaches, however, 
often 
consider a localized perspective of lane change decision-making, choosing actions that safely achieve a certain local gain~\cite{naranjo2008lane, ulbrich2015towards} and at times coordinating with local vehicles~\cite{talebpour2015modeling, wang2015game}. 
In contrast, this paper aims to examine whether longer horizon communication signals can further augment the mobility of the system.


\subsection{Lane change model: MOBIL} \label{sec:mobil}

Several prominent lane change models have been developed within the transportation community. In this study, we focus primarily on the ``Minimizing Overall Braking Induced by Lane change'' model, or MOBIL~\cite{kesting2007general, treiber2009modeling}. MOBIL is a lane change model that balances speed gain incentives for ego and neighboring vehicles when choosing a desired lane. Let $g_{v_\text{ego}}$ and $g_{v_\text{neighbors}}$ be the incentive of a given lane change from the perspective of the ego and neighboring vehicles, respectively, defined as:
\begin{equation}
    g_{v_\text{ego}} = \tilde{a}_c - a_c
\end{equation}
\begin{equation}
    g_{v_\text{neighbors}} = (\tilde{a}_n - a_n) + (\tilde{a}_o - a_o)
\end{equation}
where $\tilde{a}_i$ and $a_i$ is the desired acceleration of vehicle $i$ if a lane change does and does not occur, respectively, $c$ denotes the ego vehicle considering a lane change, and $n$ and $o$ denote the following vehicles in the target and current lanes, respectively (see Figure~\ref{fig:intro}). Following this model, a lane change is performed if the weighted combination of the above two incentives exceed a switching threshold $\Delta a_\text{th}$, or:
\begin{equation}\label{eqn:incentive}
    g_{v_\text{ego}} + p \cdot g_{v_\text{neighbors}} > \Delta a_\text{th}
\end{equation}
where $p$ is a politeness factor denoting the relevance of the incentive for neighboring vehicles. Furthermore, safety is enforced prior to any decision through a criterion that ensures that the deceleration of the successor $\tilde{a}_n$ in the target lane does not exceed a given safe limit $b_\text{safe}$, i.e.
\begin{equation}
    \tilde{a}_n \geq b_\text{safe}
\end{equation}

The role of the ``politeness'' component of this model on the decision-making of future AVs has been the topic of several studies. In the works of~\cite{hu2012scheduling}~and~\cite{khan2014analyzing}, for instance, the authors find that assigning large values of $p$ to AVs in mixed-autonomy settings can provide improvements to both traffic efficiency and safety.
As we demonstrate in Section~\ref{sec:results}, however, these improvements often come at the cost of the AVs themselves, which by definition restrict their performance to improve the performance of the collective. This restriction is an undesirable property from the perspective of the passengers of the AVs. The following section introduces an egoistic non-local speed gain incentive that helps maintain macroscopic traffic flow improvements while particularly benefiting the AVs.


\section{Method} \label{sec:method}

In this section, we introduce \emph{\methodName}, a method for decentralized lane change control that exploits information from relevant features of downstream traffic. We begin by introducing the features of traffic deemed relevant for non-local sensing, and then present an incentive-based framework through which these features can be exploited to improve the performance of various lane change decisions.

\subsection{Localizing sources of instabilities} \label{sec:localizing-incentives}

In this paper, we are interested in studying the role of CAVs in subverting the onset of instabilities brought about by traffic bottlenecks. In particular, we consider a class of slow-moving or stopping bottlenecks (Figure~\ref{fig:discretionary_a}) in which a vehicle (in green) consistently operates below the free-flow speed of traffic. These incidents may result from partial lane-closures, vehicle collisions, or other slow-moving vehicles and are often accompanied by a platoon of vehicles unable or unwilling to navigate to adjacent lanes.

For this study, we define incidents of this type by the tuple: $(x_h,\ x_t,\ v_\text{avg})$, where $x_h$ is the position of the jam head, $x_t$ is the position of the jam tail, and $v_\text{avg} \in \mathbb{R}^{n_l}$ is a vector of the average speed of vehicle near the jam tail across all $n_l$ lanes\footnote{In simulation, we compute these speeds as the average speed of vehicle $\pm 50$ m from the jam tail in each lane.}. Furthermore, we assume that tuples of this form are available for any incident existing within a network. Methods through which such data can be obtained exists in the literature including the following~\cite{weil1998traffic,bauza2010road}. In real-world settings, such signals are not readily available or perfectly reliable. Therefore, in Section~\ref{sec:results-detection} we explore the effects of 
data discrepancies 
on the controller designed.

\subsection{Adapting long-horizon incentives} \label{sec:model-discretionary}

How can CAVs exploit signals provided by downstream incidents? In this study, we consider two insights for doing so in discretionary and mandatory lane change settings. Beginning with discretionary settings, we posit that CAVs nearing an incident can reduce delays both for themselves and neighboring vehicles by exiting the lane early in anticipation of the bottleneck. In addition to subverting additional personal delays, these early exits prevent the vehicle from executing aggressive lane changes near the pockets of dense traffic that emerge by the incident. As a result, these actions are less likely to disturb the system and produce additional delays for other vehicles.


To introduce the above behavior to the decision-making of CAVs, we seek to develop a model that generalizes the incentive criterion proposed in Equation~\eqref{eqn:incentive}.
To include the long-horizon criterion, we introduce an acceleration gain component based on the event dynamics. Downstream acceleration gains are computed with respect to virtual ``vehicles'' placed at the incident tail 
(Figure~\ref{fig:discretionary_b}). The 
states
of these 
vehicles are assigned based on 
the incident tail, and provided to a car-following model
$f(\cdot)$ as in~\cite{kesting2007general} to compute 
accelerations:
\begin{equation}
a_{d} = f(h_{d},  v_\alpha, v_{\text{avg},o}), \ \ \
\tilde{a}_{d} = f(h_{d}, v_\alpha, v_{\text{avg},n}) 
\end{equation}
where $a_d$, $\tilde{a}_d$ are the accelerations with respect to the downstream event in the current and target lanes, respectively, $h_d = x_t - x_\alpha$ is the distance to the leading virtual vehicle, $x_\alpha$, $v_\alpha$ are the ego vehicle position and speed, and $v_{\text{avg},o}$, $v_{\text{avg},n}$ are the average speeds near the jam tail in the original and new lanes, respectively. Acceleration gains with respect to the downstream event, $g_{v_\text{downstream}}$, can be computed as follows:
\begin{equation}
g_{v_\text{downstream}} = \tilde{a}_{d} - a_{d}
\end{equation}

The generalized incentive criterion in discretionary lane change settings using downstream gain is then:
\begin{equation}
\lambda_s \cdot g_{v_\text{ego}} + \lambda_p \cdot g_{v_\text{neighbors}} + \lambda_d \cdot g_{v_\text{downstream}} > \Delta a_\text{th}
\end{equation}
where $\lambda_s$ is a selfishness factor, $\lambda_p$ is a politeness factor,
and $\lambda_d$ is the downstream event factor.

\subsection{Mandatory Incentives} \label{sec:model-mandatory}

Integrating discretionary and mandatory models can provide additional flexibility for making lane change decisions. To include mandatory lane changes, we further generalize the incentive criterion and introduce a mandatory gain. This gain follows the insight that as an incident nears the turning point to a targeted route, e.g., an off-ramp, a CAV must accept earlier delays incurred by the incident so as not to miss the route or further disrupt the flow of vehicles behind it.

We begin by introducing an incentive structure through which mandatory lane changes can be enforced. To do so, we place virtual stopped vehicles on the lanes in which desired routes cannot be achieved. As shown in Figure~\ref{fig:mandatory}, positions of the stopped virtual vehicles are determined based on the distance to the turning point (in this case, an off-ramp) and the offset from the desired lane. The virtual vehicle is placed at the turning point if the offset is $1$ (lane next to the desired lane). For each additional offset, we shift the virtual vehicles by a safe lane changing distance ($x_\text{safe}$) to provide sufficient space for additional lane changes. Mathematically, the distance to the virtual vehicle $h_{m,l}$ in lane $l$ is:
\begin{equation}
    h_{m,l} = \begin{cases}
        x_\text{end} - x_\alpha - (l - l_t - 1) x_\text{safe}, & \text{if } l \neq l_t \\
        \infty, & \text{if } l = l_t
    \end{cases}
    \label{eq:mandatory-virtual-vehicles}
\end{equation}
where $x_\text{end}$ is the position of the turning point and $l_t$ is the lane index required by the turning point.

Following this definition of distance to virtual vehicle, gains with respect to the virtual vehicles is:
\begin{equation}
g_{v_\text{mandatory}} = \tilde{a}_{m} - a_{m} 
\end{equation}
where the acceleration components can once again be commuted given a car-following model as follows:
\begin{equation}
a_{m} = f(h_{m,o},  v_\alpha, 0), \ \ \
\tilde{a}_{m} = f(h_{m,n}, v_\alpha, 0)
\end{equation}

The generalized incentive criterion using downstream and mandatory gains is then:
\begin{equation}
\lambda_s \cdot g_{v_\text{ego}} + \lambda_p \cdot g_{v_\text{neighbors}} + \lambda_d \cdot g_{v_\text{downstream}} + \lambda_m \cdot g_{v_\text{mandatory}} > \Delta a_\text{th}
\end{equation}
where $\lambda_m$ is the mandatory lane change factor.

\begin{figure}
\centering
	\includegraphics[width=.3\textwidth]{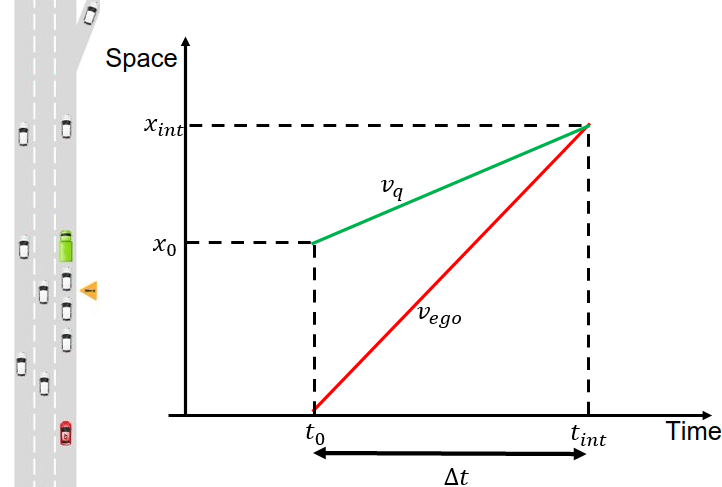}
	\caption{Overtaking Intersection Point}
	\vspace{-.5em}
	\label{fig:int_point}
\end{figure}

\begin{figure*}
    \newcommand{\laneWidth}{0.25}
    \newcommand{\separatorWidth}{5}
    \newcommand{\separatorDx}{5.25}
    \newcommand{\separatorDy}{0.125}

    \centering
    \begin{subfigure}[b]{0.48\textwidth}
        \begin{tikzpicture}
            \draw[] (\linewidth, -4.5*\laneWidth) -- (\linewidth, -5.5*\laneWidth);
            \draw[] (0, -5.5*\laneWidth) -- (0, -4.5*\laneWidth);
            \draw[] (0, -5*\laneWidth) -- (0.415*\linewidth, -5*\laneWidth);
            \draw[] (0.585*\linewidth, -5*\laneWidth) -- (\linewidth, -5*\laneWidth);
            \node () at (0.5*\linewidth, -5*\laneWidth) {\footnotesize $2$,$000$ m};

            \filldraw [fill=black!20, draw=none] (0,0) rectangle (\linewidth,3*\laneWidth);
            \draw[]       (0, 0)            -- (\linewidth, 0);
            \draw[dashed] (0, \laneWidth)   -- (\linewidth, \laneWidth);
            \draw[dashed] (0, 2*\laneWidth) -- (\linewidth, 2*\laneWidth);
            \draw[]       (0, 3*\laneWidth) -- (\linewidth, 3*\laneWidth);
            
            \draw[white, line width=\separatorWidth] 
            (0.5*\linewidth - \separatorDx, -\separatorDy) --
            (0.5*\linewidth + \separatorDx, 3*\laneWidth + \separatorDy);
            \draw[] 
            (0.5*\linewidth - \separatorDx - 0.5 * \separatorWidth, -\separatorDy) --
            (0.5*\linewidth + \separatorDx - 0.5 * \separatorWidth, 3*\laneWidth + \separatorDy);
            \draw[] 
            (0.5*\linewidth - \separatorDx + 0.5 * \separatorWidth, -\separatorDy) --
            (0.5*\linewidth + \separatorDx + 0.5 * \separatorWidth, 3*\laneWidth + \separatorDy);

            \draw[]
            (0.975*\linewidth, 0.5*\laneWidth) -- 
            (0.975*\linewidth, 6.1*\laneWidth);
            \draw[]
            (0.95*\linewidth, 1.5*\laneWidth) -- 
            (0.95*\linewidth, 4.5*\laneWidth);
            \draw[]
            (0.925*\linewidth, 2.5*\laneWidth) -- 
            (0.925*\linewidth, 4.5*\laneWidth);
            \node[anchor=east] () at (0.97*\linewidth, 5.4*\laneWidth) {\footnotesize Fast lanes};
            \node[anchor=east] () at (0.995*\linewidth, 7.00*\laneWidth) {\footnotesize Slow (incident) lane};

            \node[inner sep=0pt] () at (0.75,0.5*\laneWidth)
            {\includegraphics[height=.02\textwidth]{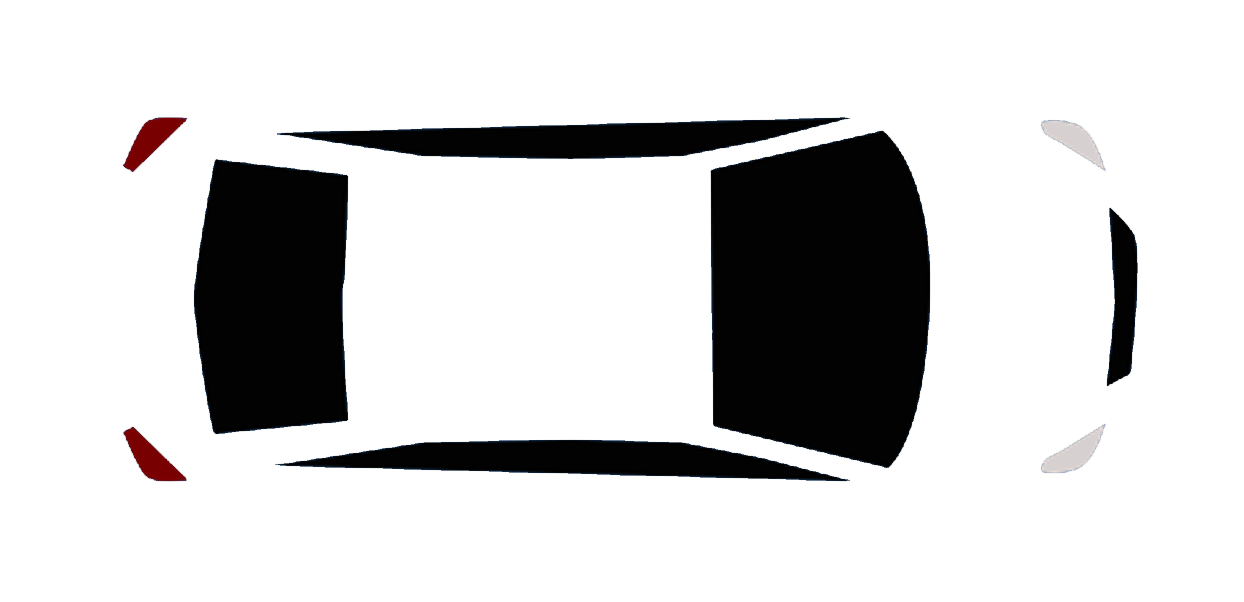}};
            \node[inner sep=0pt] () at (0.5,1.5*\laneWidth)
            {\includegraphics[height=.02\textwidth]{figures/car.png}};
            \node[inner sep=0pt] () at (1.75,0.5*\laneWidth)
            {\includegraphics[height=.02\textwidth]{figures/car.png}};
            \node[inner sep=0pt] () at (2.5,1.5*\laneWidth)
            {\includegraphics[height=.02\textwidth]{figures/car.png}};
            \node[inner sep=0pt] () at (1.0,2.5*\laneWidth)
            {\includegraphics[height=.02\textwidth]{figures/car.png}};
            \node[inner sep=0pt] () at (2.25,2.5*\laneWidth)
            {\includegraphics[height=.02\textwidth]{figures/car.png}};
            \node[inner sep=0pt] () at (3.5,2.5*\laneWidth)
            {\includegraphics[height=.02\textwidth]{figures/car.png}};
            \node[inner sep=0pt] () at (5.25,2.5*\laneWidth)
            {\includegraphics[height=.02\textwidth]{figures/car.png}};
            \node[inner sep=0pt] () at (6.25,2.5*\laneWidth)
            {\includegraphics[height=.02\textwidth]{figures/car.png}};
            \node[inner sep=0pt] () at (7.0,1.5*\laneWidth)
            {\includegraphics[height=.02\textwidth]{figures/car.png}};
            \node[inner sep=0pt] () at (8.0,0.5*\laneWidth)
            {\includegraphics[height=.02\textwidth]{figures/car.png}};

            \node[inner sep=0pt] () at (3.25,0.5*\laneWidth)
            {\includegraphics[height=.02\textwidth]{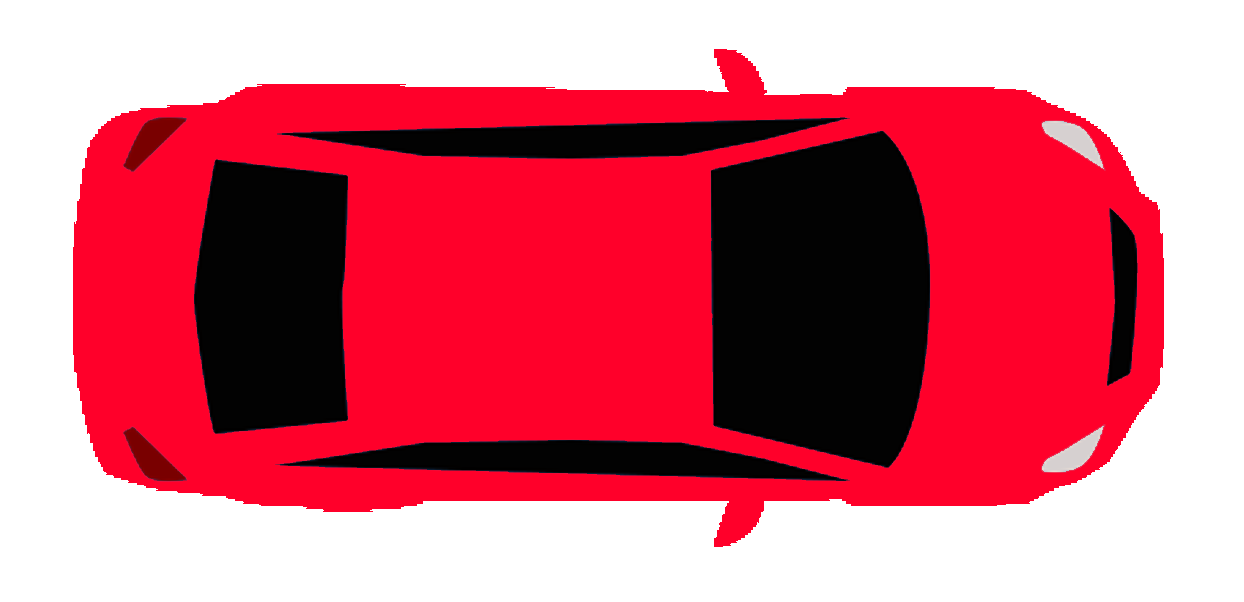}};
            \node[inner sep=0pt] () at (1.5,1.5*\laneWidth)
            {\includegraphics[height=.02\textwidth]{figures/smartcar.png}};
            \node[inner sep=0pt] () at (5.5,1.5*\laneWidth)
            {\includegraphics[height=.02\textwidth]{figures/smartcar.png}};
            \node[inner sep=0pt] () at (7.5,2.5*\laneWidth)
            {\includegraphics[height=.02\textwidth]{figures/smartcar.png}};

            \node[inner sep=0pt] () at (7,0.5*\laneWidth)
            {\includegraphics[height=.02\textwidth]{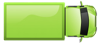}};
            \node[inner sep=0pt] () at (6.6,0.5*\laneWidth)
            {\includegraphics[height=.02\textwidth]{figures/car.png}};
            \node[inner sep=0pt] () at (6.2,0.5*\laneWidth)
            {\includegraphics[height=.02\textwidth]{figures/car.png}};
            \node[inner sep=0pt] () at (5.7,0.5*\laneWidth)
            {\includegraphics[height=.02\textwidth]{figures/car.png}};
            \node[inner sep=0pt] () at (5.0,0.5*\laneWidth)
            {\includegraphics[height=.02\textwidth]{figures/car.png}};
        \end{tikzpicture}
        \caption{Discretionary lane changes}
        \label{fig:setup-1}
    \end{subfigure}
    \hfill
    \begin{subfigure}[b]{0.48\textwidth}
        \begin{tikzpicture}
            \draw[] (\linewidth, -5.5*\laneWidth) -- (\linewidth, -4.5*\laneWidth);
            \draw[] (0.8*\linewidth, -5.5*\laneWidth) -- (0.8*\linewidth, -4.5*\laneWidth);
            \draw[] (0, -5.5*\laneWidth) -- (0, -4.5*\laneWidth);
            \draw[] (0, -5*\laneWidth) -- (0.325*\linewidth, -5*\laneWidth);
            \draw[] (0.475*\linewidth, -5*\laneWidth) -- (0.84*\linewidth, -5*\laneWidth);
            \draw[] (0.96*\linewidth, -5*\laneWidth) -- (\linewidth, -5*\laneWidth);
            \node () at (0.4*\linewidth, -5*\laneWidth) {\footnotesize $1$,$900$ m};
            \node () at (0.9*\linewidth, -5*\laneWidth) {\footnotesize $100$ m};

            \filldraw [fill=black!20, draw=none] (0,0) rectangle (\linewidth,3*\laneWidth);
            \draw[]       (0, 0)              -- (0.76*\linewidth, 0);
            \draw[black!20, line width=2]       (0.76*\linewidth, 0) -- (0.8*\linewidth, 0);
            \draw[]       (0.8*\linewidth, 0) -- (\linewidth, 0);
            \draw[dashed] (0, \laneWidth)     -- (\linewidth, \laneWidth);
            \draw[dashed] (0, 2*\laneWidth)   -- (\linewidth, 2*\laneWidth);
            \draw[]       (0, 3*\laneWidth)   -- (\linewidth, 3*\laneWidth);
            
            \draw[fill=black!20, draw=none] 
            (0.80*\linewidth - 0.5*\laneWidth, 0) -- 
            (0.94*\linewidth - 0.5*\laneWidth, -3*\laneWidth) --
            (0.90*\linewidth - 0.5*\laneWidth, -3*\laneWidth) --
            (0.76*\linewidth - 0.5*\laneWidth, 0);
            \draw[] 
            (0.76*\linewidth - 0.5*\laneWidth, 0) -- 
            (0.90*\linewidth - 0.5*\laneWidth, -3*\laneWidth);
            \draw[] 
            (0.80*\linewidth - 0.5*\laneWidth, 0) -- 
            (0.94*\linewidth - 0.5*\laneWidth, -3*\laneWidth);

            \node[inner sep=0pt] () at (0.5,1.5*\laneWidth)
            {\includegraphics[height=.02\textwidth]{figures/car.png}};
            \node[inner sep=0pt] () at (2.25,2.5*\laneWidth)
            {\includegraphics[height=.02\textwidth]{figures/car.png}};
            \node[inner sep=0pt] () at (3.5,2.5*\laneWidth)
            {\includegraphics[height=.02\textwidth]{figures/car.png}};
            \node[inner sep=0pt] () at (5.25,2.5*\laneWidth)
            {\includegraphics[height=.02\textwidth]{figures/car.png}};
            \node[inner sep=0pt] () at (6.25,2.5*\laneWidth)
            {\includegraphics[height=.02\textwidth]{figures/car.png}};
            \node[inner sep=0pt] () at (7.0,1.5*\laneWidth)
            {\includegraphics[height=.02\textwidth]{figures/car.png}};
            \node[inner sep=0pt] () at (8.0,0.5*\laneWidth)
            {\includegraphics[height=.02\textwidth]{figures/car.png}};
            \node[inner sep=0pt] () at (1.5,1.5*\laneWidth)
            {\includegraphics[height=.02\textwidth]{figures/car.png}};
            \node[inner sep=0pt] () at (7.5,2.5*\laneWidth)
            {\includegraphics[height=.02\textwidth]{figures/car.png}};

            \node[inner sep=0pt] () at (1.75,0.5*\laneWidth)
            {\includegraphics[height=.02\textwidth]{figures/smartcar.png}};

            \node[inner sep=0pt] () at (5,0.5*\laneWidth)
            {\includegraphics[height=.02\textwidth]{figures/truck.png}};
            \node[inner sep=0pt] () at (4.6,0.5*\laneWidth)
            {\includegraphics[height=.02\textwidth]{figures/car.png}};
            \node[inner sep=0pt] () at (4.2,0.5*\laneWidth)
            {\includegraphics[height=.02\textwidth]{figures/car.png}};
            \node[inner sep=0pt] () at (3.7,0.5*\laneWidth)
            {\includegraphics[height=.02\textwidth]{figures/car.png}};
            \node[inner sep=0pt] () at (3.0,0.5*\laneWidth)
            {\includegraphics[height=.02\textwidth]{figures/car.png}};

            \draw[white, line width=\separatorWidth] 
            (0.12*\linewidth - \separatorDx, -\separatorDy) --
            (0.12*\linewidth + \separatorDx, 3*\laneWidth + \separatorDy);
            \draw[] 
            (0.12*\linewidth - \separatorDx - 0.5 * \separatorWidth, -\separatorDy) --
            (0.12*\linewidth + \separatorDx - 0.5 * \separatorWidth, 3*\laneWidth + \separatorDy);
            \draw[] 
            (0.12*\linewidth - \separatorDx + 0.5 * \separatorWidth, -\separatorDy) --
            (0.12*\linewidth + \separatorDx + 0.5 * \separatorWidth, 3*\laneWidth + \separatorDy);

            \draw[->,red] 
            (1.75, 0.5*\laneWidth) -- 
            (2.10, 1.5*\laneWidth) --
            (6.30, 1.5*\laneWidth) --
            (6.7, 0.5*\laneWidth);
        \end{tikzpicture}
        \caption{Mandatory overtaking}
        \label{fig:setup-2}
     \end{subfigure}
    \caption{An illustration of the numerical experiments explored within this paper. \textbf{Left:} An incident vehicle (in green) reduces the throughput of a single lane, and AVs (in red) perform lane changes to avoid the subsequent congestion. \textbf{Right:} An off-ramp is introduced, and AVs are tasks with performed lane change actions that reduce the time needed to reach the exit while maximizing the success rate.}
    \vspace{-.5em}
    \label{fig:experimental-setups}
\end{figure*}

With the mandatory incentive structure defined, we next introduce a mechanism to exploit incident information to reduce delays and improve the success rate of routed CAVs. To do so, we realign the position of the virtual vehicles to increase the urgency of the mandatory lane change incentive whenever deemed necessary, as stated at the start of this subsection.
Figure~\ref{fig:mandatory} shows a slow vehicle scenario where a queue is built up behind the slow vehicle. A CAV that wants to take the next exit is approaching this event and must choose whether to overtake the slow queue. To determine if this is possible, we first estimate the point where the CAV and jam head intersect if the CAV attempts to overtake (Figure~\ref{fig:int_point}). The intersection point can be determined based the initial distance between the CAV and jam head, $x_h - x_\alpha$, and estimated speed near the jam in both the incident $v_{\text{avg},o}$ and adjacent $v_{\text{avg},n}$ lanes as follows:
\begin{equation}
x_\text{int} = x_\alpha + v_{\text{avg},n} \left( \frac{x_\alpha - x_h}{v_{\text{avg},o} - v_{\text{avg},n}} \right)
\end{equation}

If the intersection point is not sufficiently far from the exit ($x_\text{int} + x_\text{safe} > x_\text{m}$), we realign the virtual vehicles to the jam tail. This forces CAVs to stay behind the slow queue when overtaking is unlikely. 
Otherwise,
we maintain the original vehicle positions,
allow overtaking actions. Mathematically, the updated virtual vehicle headways $h_{m,l}'$ are:
\begin{equation}
    h_{m,l}' = \begin{cases}
        h_{m,l} - x_\text{end} + x_t, & \text{if } x_\text{int} + x_\text{safe} > x_\text{m} \\
        h_{m,l}, & \text{otherwise}
    \end{cases}
\end{equation}

These new headways are assignment to the mandatory incentives $g_{v_\text{mandatory}}$ for the CAVs. In Figure~\ref{fig:mandatory}, the intersection point is not sufficiently far from the exit; therefore, we place the virtual vehicles (light blue color vehicles) near the jam tail which disincentivize the overtaking maneuver. In the following section, we explore the efficacy of the proposed model.

\section{Numerical Results} \label{sec:results}

In this section, we present numerical results for the proposed model on a number of relevant problems. Though these experiments, we aim to answer the following questions:
\begin{enumerate}
    \item Is the proposed model effective at improving the mobility of the controlled vehicles, and how does it compare to models that rely solely on local sensing?
    \item What influence, if any, does this model have on global properties of traffic?
\end{enumerate}

We begin in Sec.~\ref{sec:problem-setup}~to~\ref{sec:simulations} by introducing the considered problems and simulation procedure. We then describe the performance of the proposed model within these problems in Sec.~\ref{sec:results-disc}~to~\ref{sec:results-mand}. All results 
are reported across $100$ runs 
to account for stochasticity between simulations.

\subsection{Problem setup} \label{sec:problem-setup}

Experiments are conducted on a multi-lane highway network with varying configurations, see Figure~\ref{fig:experimental-setups}. Specifically, we consider two problems designed to assess the responsiveness of the model to both discretionary and mandatory lane changing settings. These tasks are both explored for inflow rates of \{$800$, $1000$, $1200$, $1400$\} veh/hr/lane.

\subsubsection{Discretionary lane changes}

In the first of these tasks (Figure~\ref{fig:setup-1}), a slow-moving or stopping vehicle is placed on the rightmost lane of a $3$-lane, $2$ km long highway. In particular, stopping vehicles are placed $1.5$ km from the front of the network, while slow-moving vehicles are placed $100$~m and 
drive at 
$10$ m/s. 
The presence of these vehicles on the 
network produces growing regions of slow-moving traffic
that reduce global throughput.
The objective of
CAVs in this case is to 
minimize delays resulting from the incident.

\subsubsection{Mandatory overtaking}

In the second of these tasks (Figure~\ref{fig:setup-2}), an off-ramp is introduced 
$1.9$ km from the start of the edge. The off-ramp produces an additional route which is assigned to $20$\% of the entering vehicles. 
The ability for vehicles to reach the targeted off-ramp and exit the freeway,
however, is hindered by 
the presence of a slow-moving vehicle (in this case operating at $5$ m/s). 
The objective of CAVs in this setting is to ensure successful maneuvering to the off-ramp while not restricting 
global mobility.

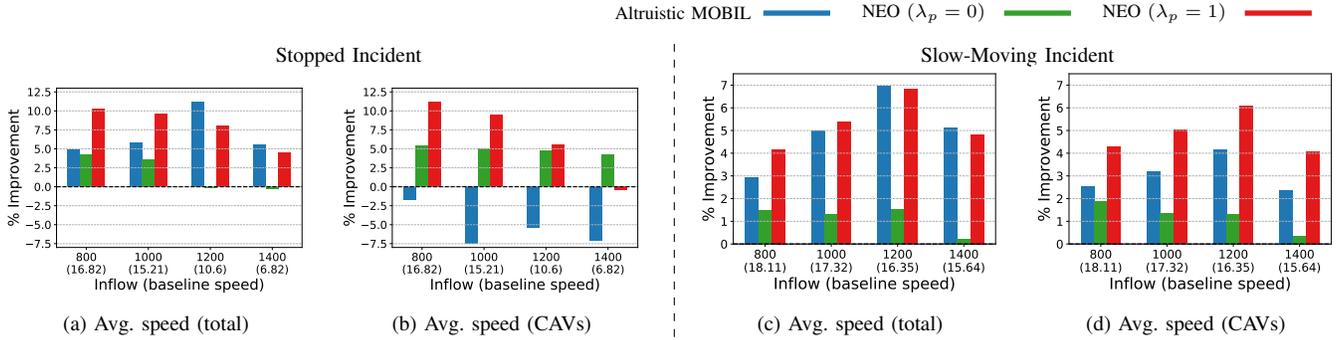
\begin{figure*}
\begin{subfigure}[b]{\textwidth}
    \begin{tikzpicture}
        \node () at (0, -0.2) {\textcolor{white}{.}};
        \node () at (0.5*\linewidth, 0) {\scriptsize Altruistic MOBIL};
        \draw[color=color2, line width=2] (0.56*\linewidth, 0) -- (0.61*\linewidth, 0);
        \node () at (0.68*\linewidth, 0) {\scriptsize \methodName $(\lambda_p = 0)$};
        \draw[color=color4, line width=2] (0.74*\linewidth, 0) -- (0.79*\linewidth, 0);
        \node () at (0.86*\linewidth, 0) {\scriptsize \methodName $(\lambda_p = 1)$};
        \draw[color=color6, line width=2] (0.92*\linewidth, 0) -- (0.97*\linewidth, 0);
    \end{tikzpicture}
\end{subfigure}\\
\begin{subfigure}[b]{\textwidth}
    \begin{tikzpicture}
        \node () at (0, 0) {\textcolor{white}{.}};
        \node () at (0.25*\linewidth, 0) {\footnotesize Stopped Incident};
        \node () at (0.75*\linewidth, 0) {\footnotesize Slow-Moving Incident};
    \end{tikzpicture}
\end{subfigure}\\[-12pt]
\begin{subfigure}[b]{.23\textwidth}
	\includegraphics[width=\textwidth]{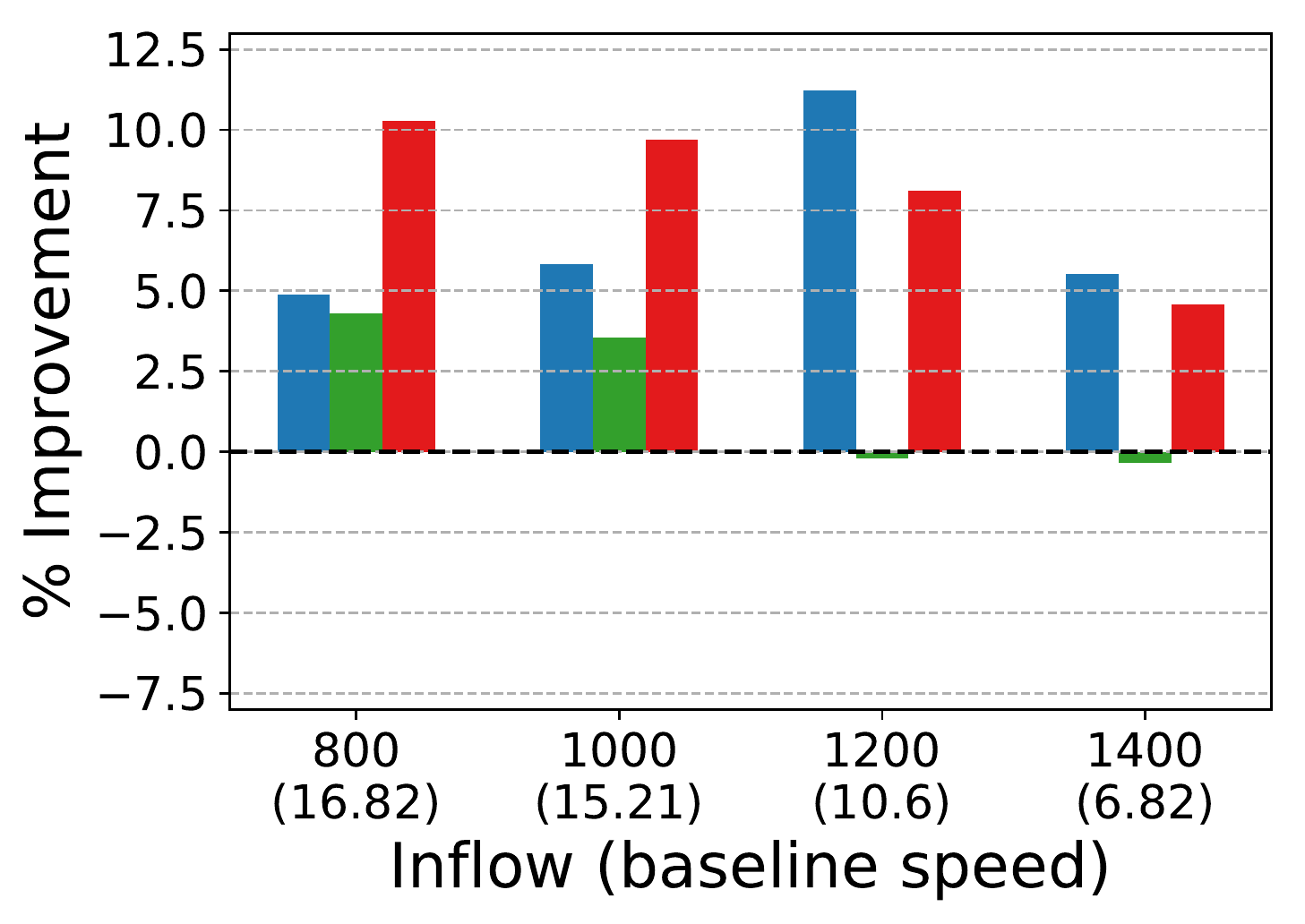}
	\caption{Avg. speed (total)}
\end{subfigure}
\hfill
\begin{subfigure}[b]{.23\textwidth}
    \centering
	\includegraphics[width=\textwidth]{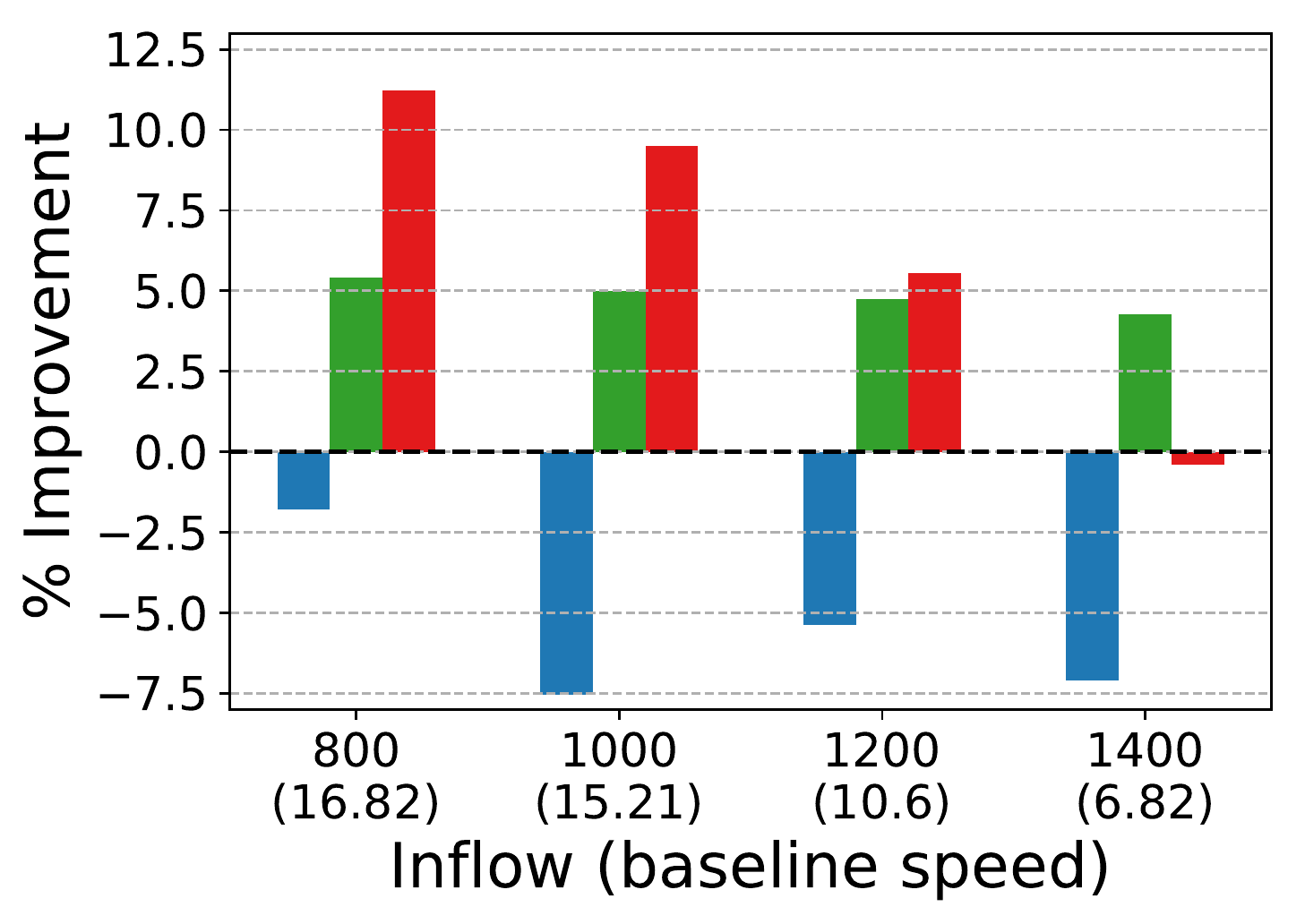}
	\caption{Avg. speed (CAVs)}
\end{subfigure}
\hfill
\begin{subfigure}[b]{.01\textwidth}
    \begin{tikzpicture}
        \draw[dashed] (0, 0) -- (0, 3.9);
    \end{tikzpicture}
\end{subfigure}
\begin{subfigure}[b]{.23\textwidth}
	\includegraphics[width=\textwidth]{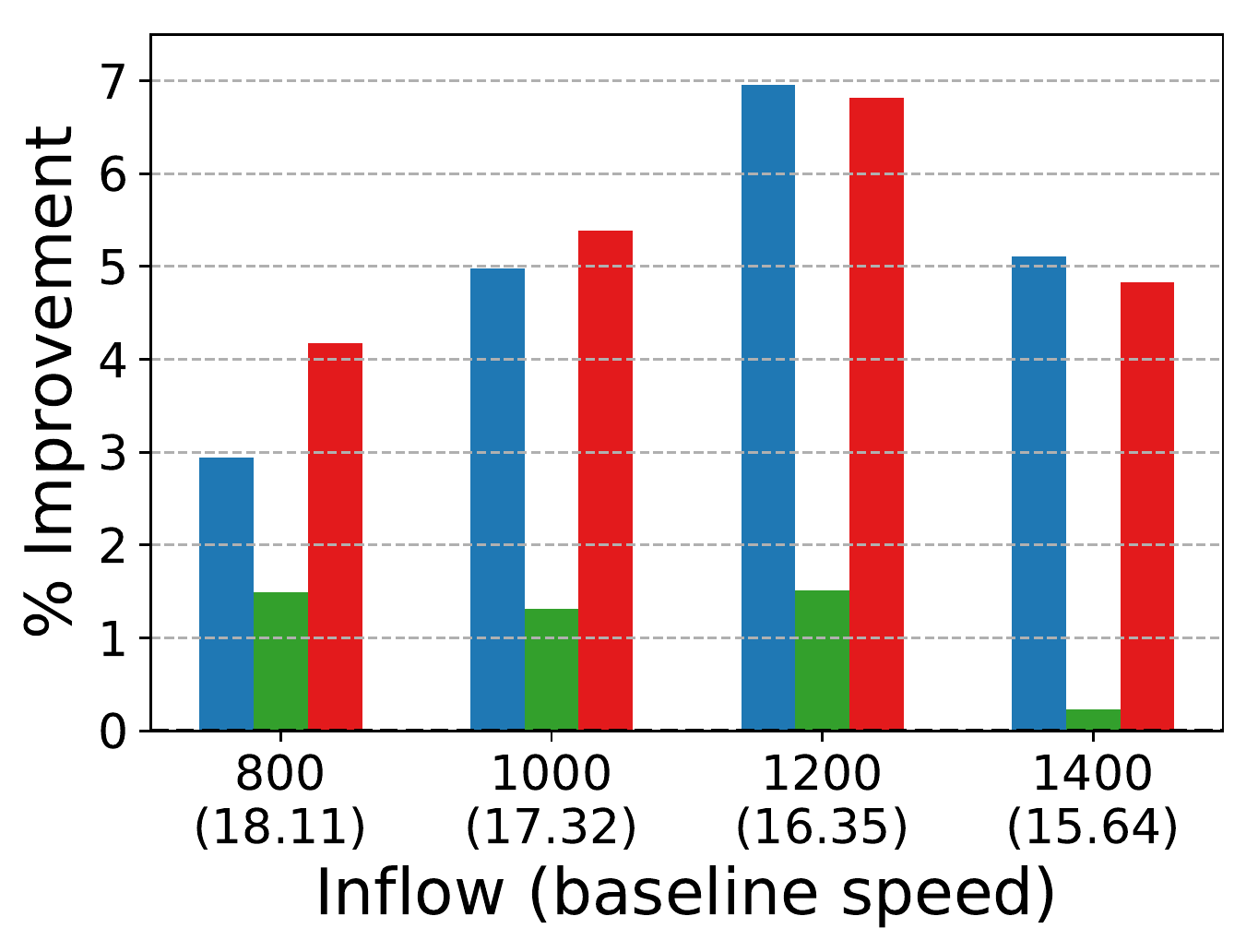}
	\caption{Avg. speed (total)}
\end{subfigure}
\hfill
\begin{subfigure}[b]{.23\textwidth}
    \centering
	\includegraphics[width=\textwidth]{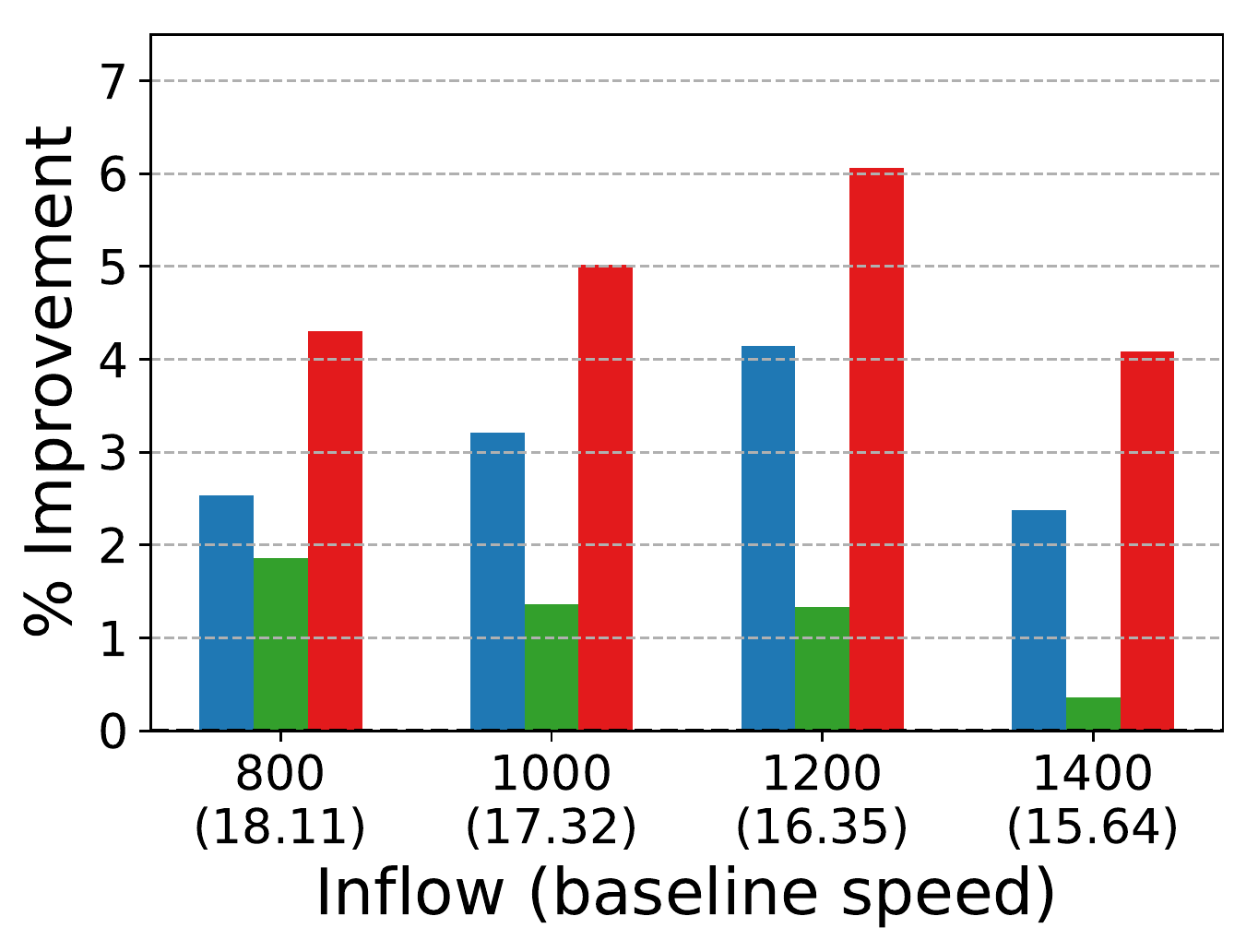}
	\caption{Avg. speed (CAVs)}
\end{subfigure}
\caption{Effect of choice of model parameters in the presence of a slow-moving and stopping incident. For both slow-moving and stopping incidents, our controller largely succeeds in both improving the global and CAV-level mobility of vehicles within the network. Moreover, while the ``politeness'' incentives in isolation (blue) restrict the mobility of CAVs, when combined with our discretionary incentives (red) provides on average greater improvement to mobility and more balanced benefits to the CAVs.}
\vspace{-.5em}
\label{fig:discretionary-param-effects}
\end{figure*}

\subsection{Human-driver dynamics} \label{sec:human-models}

The car-following behaviors of individual vehicles are defined by the \emph{Intelligent Driver Model} (IDM)~\cite{treiber2000congested}.
Through this model, the acceleration for a  vehicle $\alpha$ is defined by its bumper-to-bumper headway $h_\alpha$, velocity $v_\alpha$, and relative velocity with the preceding vehicle $\Delta v_\alpha$ as:
\begin{equation}
\resizebox{.42\textwidth}{!}{
$f_\text{IDM} (h_\alpha, v_\alpha, \Delta v_\alpha) = a \left[ 1 - \left( \frac{v_\alpha}{v_0} \right)^\delta - \left( \frac{s^*(v_\alpha, \Delta v_\alpha)}{h_\alpha} \right)^2 \right] + \epsilon$%
}
\end{equation}
where $\epsilon$ is an exogenous noise term to mimic stochasticity in human driving, $s^*$ is the desired headway denoted by:
\begin{equation}
s^*(v_\alpha, \Delta v_\alpha) = s_0 + \text{max}\left(0, v_\alpha T + \frac{v_\alpha \Delta v_\alpha}{2 \sqrt{ab}} \right)
\end{equation}
and $h_0$, $v_0$, $T$, $\delta$, $a$, $b$ are given parameters provided below.

\begin{center}
    \small
    \begin{tabular}{|l|c|c|c|c|c|c|c|}
    \hline
    \textbf{Parameter} & $v_0$ & $T$ & $a$ & $b$ & $\delta$ & $s_0$ & $\epsilon$ \\ \hline
    \textbf{Value} & $20$ & $1.2$ & $1.5$ & $2.0$ & $4$ & $2$ & $\mathcal{N}(0,0.2)$ \\ \hline
    \end{tabular}
\end{center}

Next, to model the lane change behaviors of human drivers, we look to the MOBIL lane change model depicted in Section~\ref{sec:mobil},
and use an
egoistic variation of the model 
where
$p$ is set to $0$ as proposed in~\cite{kesting2007general}. 
To estimate the accelerations for the individual speed-gain incentives within this model (both for the humans and CAVs in future sections), we utilize the IDM model presented earlier in this subsection.

Finally, to model that mandatory 
behavior of humans attempting to exit the off-ramp, we implement a local variant of the behavior presented in Section~\ref{sec:model-mandatory}. Specifically, for vehicles expected to exit the off-ramp, we place virtual stopped vehicles at the off-ramp following Eq.~\eqref{eq:mandatory-virtual-vehicles}, without re-positioning the vehicles 
in response to
the incident. 
In practice, we find that this incentive is overridden at times by the selfish speed gain component, resulting in more vehicles failing to reach the off-ramp.
Therefore, to address this, we linearly anneal the selfishness coefficient for routing vehicles from its max value $1000$ m from the off-ramp to $0$ at the turning point. 
This is done
both for humans and CAVs. 

\subsection{Simulations} \label{sec:simulations}

Simulations are conducted for a step size of $0.25$ sec/step. These are run for a total of $1200$ sec, or until the slow-moving vehicles reach the end of the network, whichever comes first. To model an inflow rate $q_\text{in}$ (in veh/sec), vehicles are introduced from the start of the network everything $1/q_\text{in}$ sec driving at free-flow speeds and randomly among all lanes. Finally, to model different penetration rates $p_\text{CAV}$, we replace an entry vehicle with probability $p_\text{CAV}$ with a vehicle whose lane change actions are provided by a model of choice.

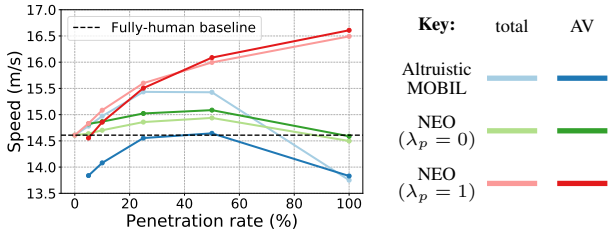
\begin{figure}
\begin{subfigure}[b]{.275\textwidth}
	\includegraphics[width=\textwidth]{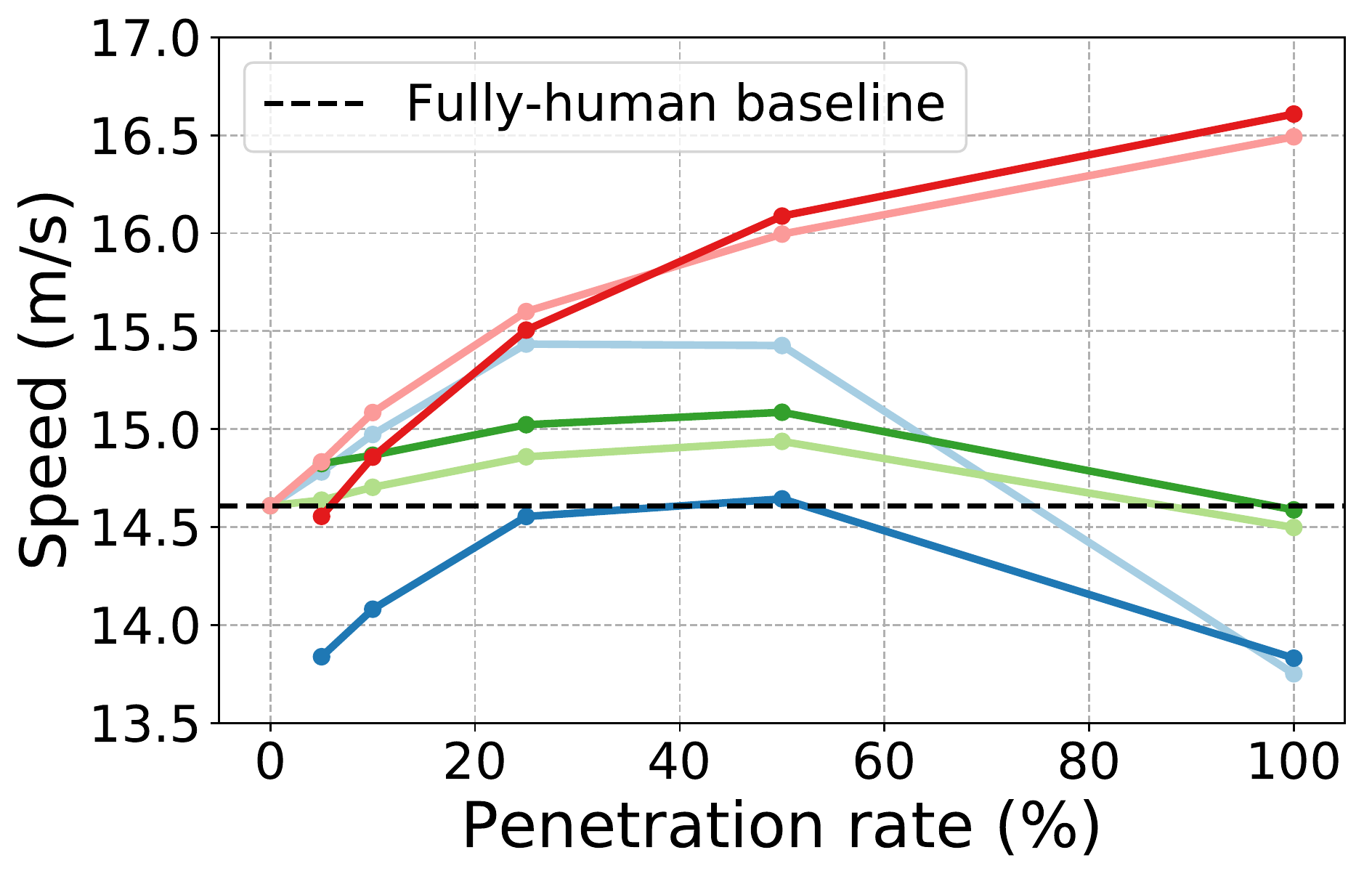}
\end{subfigure}
\
\begin{subfigure}[b]{.15\textwidth}
    \begin{tikzpicture}
        \node () at (0.*\linewidth, -2.65) {\textcolor{white}{.}};
        \draw[color=color1, line width=2] (0.3*\linewidth, -0.7) -- (0.55*\linewidth, -0.7);
        \draw[color=color2, line width=2] (0.65*\linewidth, -0.7) -- (0.9*\linewidth, -0.7);
        \draw[color=color3, line width=2] (0.3*\linewidth, -1.4) -- (0.55*\linewidth, -1.4);
        \draw[color=color4, line width=2] (0.65*\linewidth, -1.4) -- (0.9*\linewidth, -1.4);
        \draw[color=color5, line width=2] (0.3*\linewidth, -2.1) -- (0.55*\linewidth, -2.1);
        \draw[color=color6, line width=2] (0.65*\linewidth, -2.1) -- (0.9*\linewidth, -2.1);
        \node () at (0.05*\linewidth, -0.575) {\scriptsize Altruistic};
        \node () at (0.05*\linewidth, -0.825) {\scriptsize MOBIL};
        \node () at (0.05*\linewidth, -1.275) {\scriptsize \methodName};
        \node () at (0.05*\linewidth, -1.525) {\scriptsize $(\lambda_p = 0)$};
        \node () at (0.05*\linewidth, -1.975) {\scriptsize \methodName};
        \node () at (0.05*\linewidth, -2.225) {\scriptsize $(\lambda_p = 1)$};
        \node () at (0.05*\linewidth, 0) {\scriptsize \textbf{Key:}};
        \node () at (0.425*\linewidth, 0) {\scriptsize total};
        \node () at (0.775*\linewidth, 0) {\scriptsize AV};
    \end{tikzpicture}
\end{subfigure}
\caption{Effect of penetration rate on the discretionary incentives. While the existing model and fully egoistic \methodName model fail to perform well at higher penetration rates, the combination of the two 
consistently outperforms human-drivers and continues to grow.
}
\vspace{-.5em}
\label{fig:disc-penetration-effects}
\end{figure}

\subsection{Discretionary lane changes} \label{sec:results-disc}

We begin by studying the impact of the long-horizon incentive in the problem setup described in Section~\ref{sec:model-discretionary}. We compare the performance of this incentive structure against two baselines: 1) a fully human-driven baseline, and 2) an ``\benchmarkName'' model in which the politeness value $p$ for the lane change model of automated vehicles is set to $1$. This is proposed in~\cite{khan2014analyzing} as a method of improving traffic-efficiency, and can be seen as a special case of \methodName where $\lambda_d$ and $\lambda_m$ are set to $0$. For each of these controllers, we evaluate performance 
based on two metrics.
First, to assess global improvements to mobility, we use the average speeds of all drivers (both humans and CAVs) as a surrogate. 
Next, to ensure gains
are not imbalanced between humans and CAVs, 
we also compute the average speeds of CAVs.


\begin{figure}
\begin{subfigure}[b]{.22\textwidth}
	\includegraphics[height=0.72\textwidth]{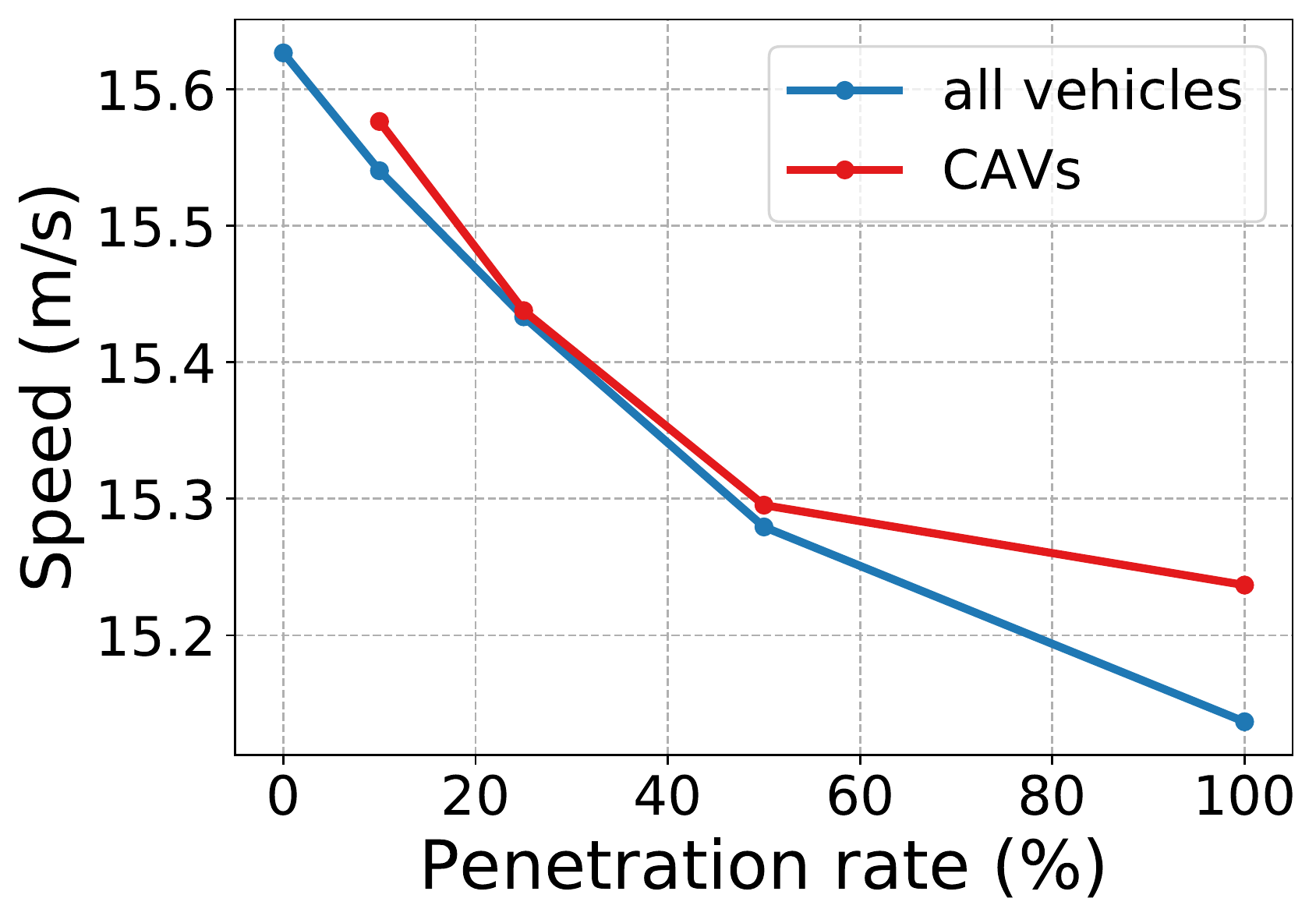}
	\caption{Average speed}
\end{subfigure}
\hfill
\begin{subfigure}[b]{.22\textwidth}
    \centering
	\includegraphics[height=0.72\textwidth]{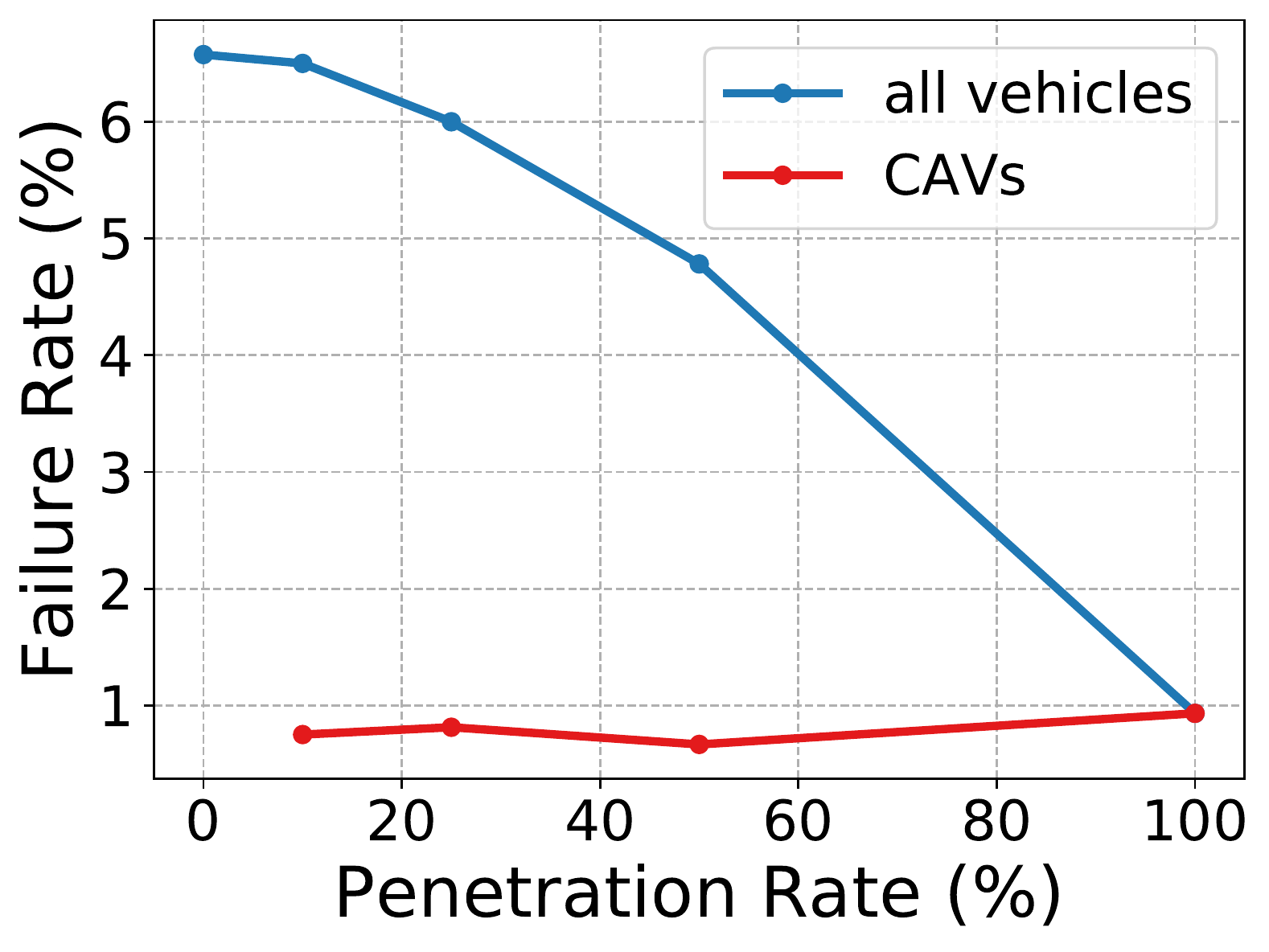}
	\caption{Off-ramp failure rate}
\end{subfigure}
\caption{Effect of penetration rate on the mandatory incentives. The CAVs consistently outperform human drivers in reaching the desired off-ramp, and do so without seriously reducing driving speeds.}
\vspace{-.5em}
\label{fig:mandatory-results}
\end{figure}


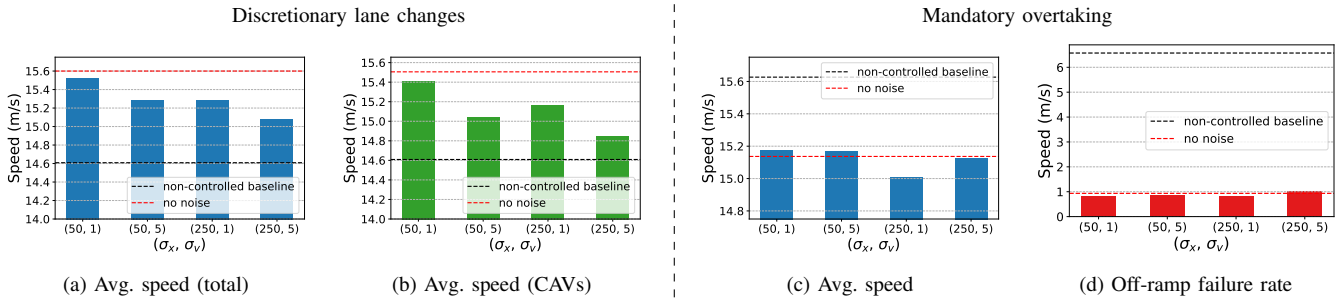
\begin{figure*}
\begin{subfigure}[b]{\textwidth}
    \begin{tikzpicture}
        \node () at (0, 0) {\textcolor{white}{.}};
        \node () at (0.25*\linewidth, 0) {\footnotesize Discretionary lane changes};
        \node () at (0.75*\linewidth, 0) {\footnotesize Mandatory overtaking};
    \end{tikzpicture}
\end{subfigure}\\[-12pt]
\begin{subfigure}[b]{.23\textwidth}
	\includegraphics[width=\textwidth]{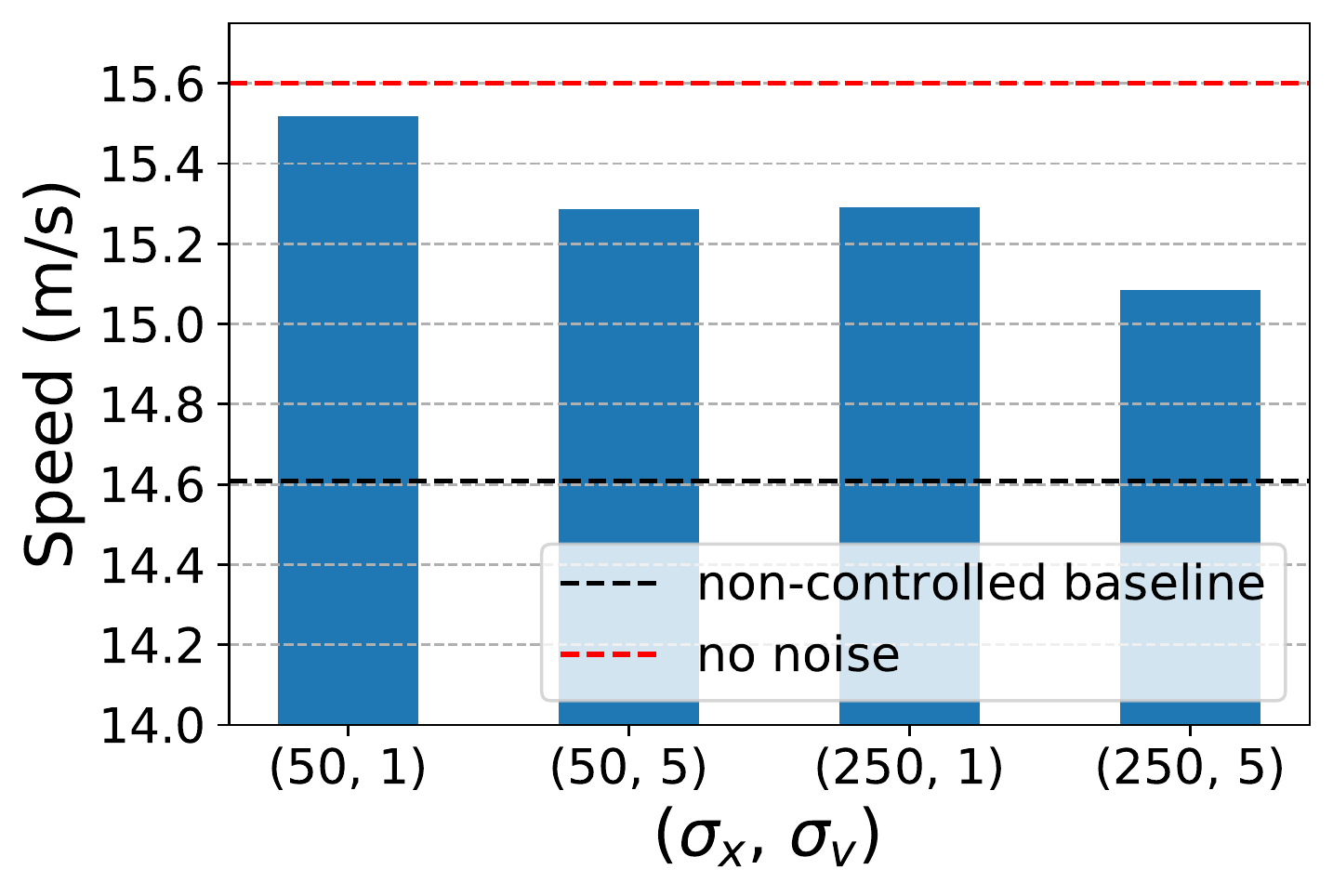}
	\caption{Avg. speed (total)}
\end{subfigure}
\hfill
\begin{subfigure}[b]{.23\textwidth}
    \centering
	\includegraphics[width=\textwidth]{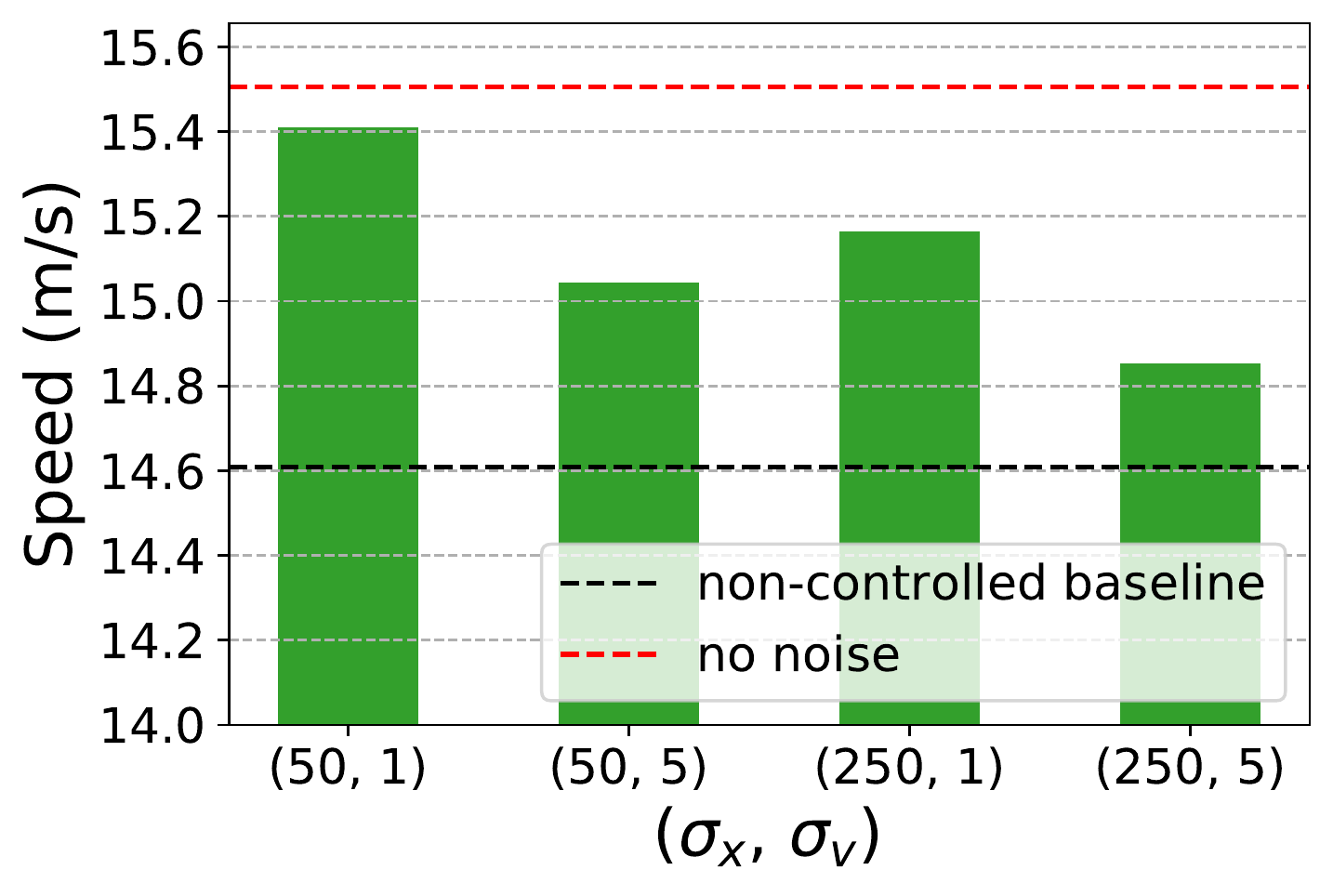}
	\caption{Avg. speed (CAVs)}
\end{subfigure}
\hfill
\begin{subfigure}[b]{.01\textwidth}
    \begin{tikzpicture}
        \draw[dashed] (0, 0) -- (0, 3.9);
    \end{tikzpicture}
\end{subfigure}
\begin{subfigure}[b]{.23\textwidth}
	\includegraphics[width=\textwidth]{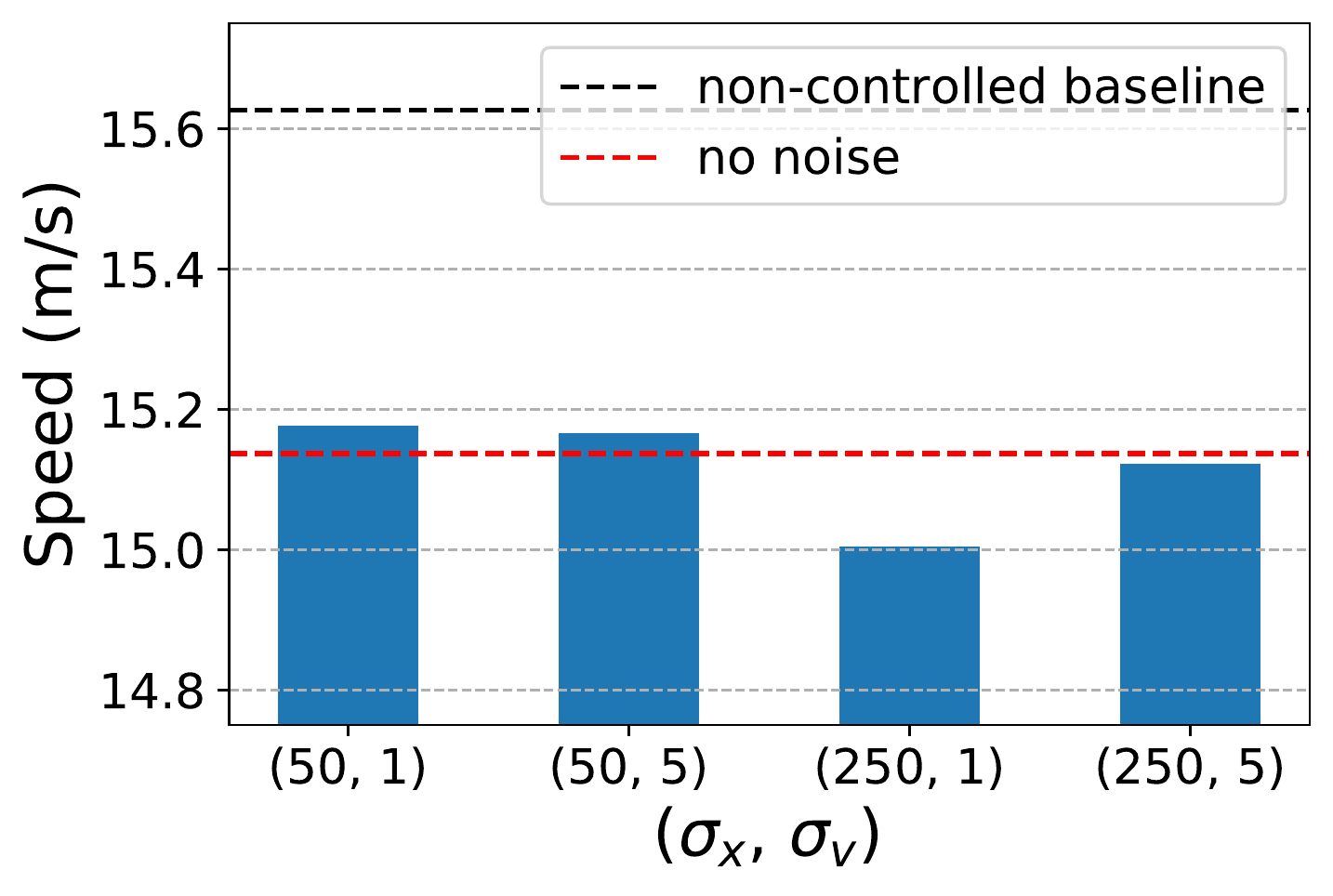}
	\caption{Avg. speed}
\end{subfigure}
\hfill
\begin{subfigure}[b]{.23\textwidth}
    \centering
	\includegraphics[width=\textwidth]{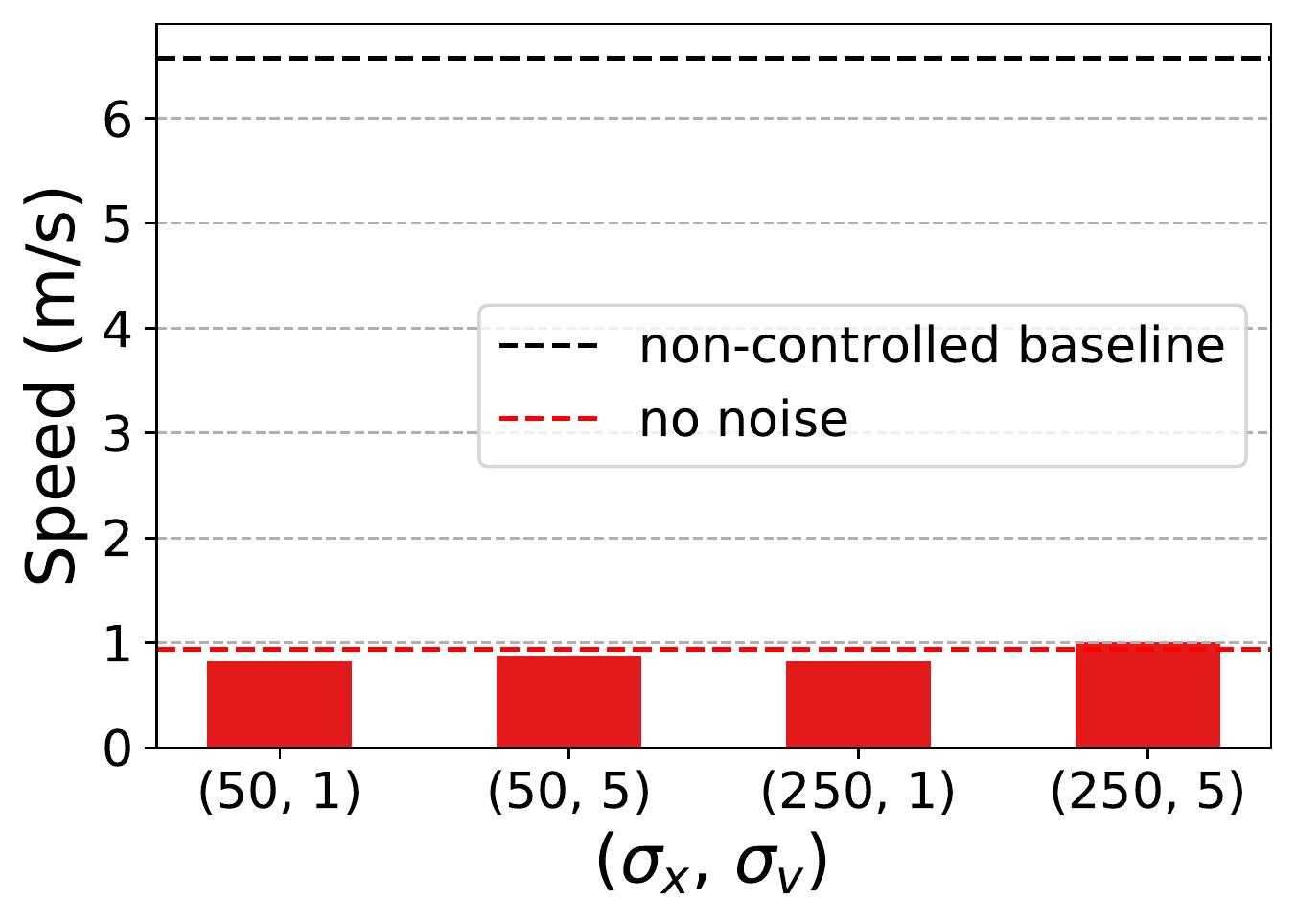}
	\caption{Off-ramp failure rate}
\end{subfigure}
\caption{The effect of incident detection errors on the performance of the model. \textbf{Left:} For the discretionary tasks, while some degradation occurs, the model continues to improve upon human-driven performances even for significant detection errors. \textbf{Right:} Interestingly enough, the addition of noise has little effect on the mandatory incentives, highlighting the benefits of simply identifying bottlenecks.}
\vspace{-.5em}
\label{fig:detection-error-results}
\end{figure*}

Figure~\ref{fig:discretionary-param-effects} depicts the gains (compared to a purely human-driven setting) of the proposed and existing models at a penetration rate of 20\%.
We highlight the following findings from these results.
\begin{itemize}
    \item \benchmarkName: While the \benchmarkName model does succeed at improving system-level mobility, it does so at the cost of the CAVs, with CAVs experiencing slower driving speeds around a stopped incident and reduced gains around a slow-moving incident. This disparity results from the CAV's reluctance to exit the incident lane when such actions may perturb adjacent lanes.
    \item \methodName ($\lambda_p = 0$): In the absence of the politeness component, the \methodName model does provide balanced gains between humans and CAVs, but experiences reduced gains as a whole. The gains occur due to performing exits from the incident lane before entering dense traffic settings. While these gains can be improved upon, it is worth noting that prematurely exiting alone provides significant gains.
    \item \methodName ($\lambda_p = 1$): Finally, when both the politeness and long-horizon components are employed, we remarkably see gains often higher than those experienced globally by the \benchmarkName model, and a more balanced distribution between humans and CAVs. This result highlights the additive benefit of the long-horizon component when introduced to existing models and motivates incorporating this incentive to the MOBIL model in particular.
\end{itemize}

Figure~\ref{fig:disc-penetration-effects} depicts the effect of CAV penetration rate on the performance of the controller, averaged across all inflows. Interestingly, we find here that while the politeness and long-horizon incentives alone fail to perform as well at higher penetration rates, the combination of the two consistently outperforms all others and continues to grow as penetration rates increase. This highlights a potential limitation to the purely long horizon incentive, with vehicles densely packing the lane adjacent to the incident, and provides further motivation for coupling this incentive with MOBIL. 

For the \methodName model in the above experiments, we use a selfishness factor $\lambda_s$ of $1$ and a long horizon factor $\lambda_d$ of $100$. The need for relatively large coefficient factors for the long horizon incentive comes from the large distances between the ego vehicle and the incident. For such distances, relative gains from the perspective of standard car-following models are minute. As a result, large coefficients are needed to ensure that these gains can trigger the decision-making threshold of the model far enough from the incident.

\subsection{Mandatory overtaking} \label{sec:results-mand}

We next study the impact of the mandatory incentives on vehicles attempting to reach the off-ramp. Unlike the previous section, we ignore the use of the politeness incentive, as the use of such incentives reduce the likelihood of vehicles performing lane changing needed to reach a given route (this is highlighted in the paper which originally introduces the MOBIL model~\cite{kesting2007general} and concurs with the finding in Section~\ref{sec:results-disc}). 
Moreover, for all vehicles in the network we consider a mandatory incentive scale of value $\lambda_m = 100$. The motivation for this value is in the previous subsection.

Figure~\ref{fig:mandatory-results} depicts the effect of CAV penetration rate on system-level mobility and the failure rate\footnote{For failure rates, we denote a mandatory lane change maneuver as a ``failure'' if the vehicle do not reach the off-ramp lane $10$ m prior to the off-ramp, at which point it is rerouted through the main highway.} of vehicles attempting to reach the off-ramp, once again averaged across all inflows. 
In terms of 
aggregate mobility, 
we see minor reductions to the average speed of vehicles. It should be noted that the slow moving vehicle is not a CAV which causes a disparity in average speeds of all vehicles and CAVs when the penetration rate is 100\%. The losses in average speed of vehicles are counterbalanced by improvements to the failure rate of routing vehicles, in which we see that CAVs succeed to exit the off-ramp 
at a far higher rate than human drivers, who do not account for the incident when choosing mandatory actions.
This reduction has little effect on the humans' ability to reach the off-ramp, with system-level reductions arising primarily from the CAVs. The mandatory incentive structure, as such, serves to benefits CAVs but does little else to other drivers.
In future studies, we hope to couple this incentive with similar ``politeness'' incentives for mandatory lane change settings, ideally providing similar additive benefits as witnessed in discretionary settings.

\subsection{Effect of incident detection errors} \label{sec:results-detection}

Finally, we evaluate the robustness of the model to errors in both the localization of the incident and the estimation of properties of traffic in adjacent lanes. To do so, we introduce Gaussian noise to the variables in Section~\ref{sec:localizing-incentives}. Specifically, for the position components ($x_t$ and $x_h$) we add noise with std dev $\sigma_x \in \{ 50, 250 \}$ m, while for the speed components ($v_\text{avg}$), we add noise with std dev $\sigma_v \in \{ 1, 5 \}$ m/s.

Figure~\ref{fig:detection-error-results} depicts the effect of incident detection errors on the performance of both discretionary and mandatory incentives. We use the same metrics as before to measure the effects on mobility for both tasks, and for simplicity, we average the results across all incident and inflow conditions. For the discretionary tasks, the addition of noise to the system results in moderate reductions in the model's performance. Notably, however, gains are still experienced for even significant detection errors, highlighting the robustness of the proposed incentive. Similar results may be noted for the mandatory incentives, where interestingly, we find that the addition of noise has little effect on performance.

\section{Conclusion} \label{sec:conclusion}

This paper explores methods for exploiting knowledge of downstream traffic bottlenecks to improve the lane change behaviors of CAVs in freeway networks. It introduces two incentives to handle such bottlenecks in discretionary and mandatory settings, subverting bottlenecks when discretionary 
actions
are required and tracking them if mandatory overtaking 
is
not possible. These incentives are demonstrated to improve the traffic-flow efficiency of several different tasks while not disproportionately hindering the CAVs. 
As a topic of ongoing work, we aim to study the sensitivity of the proposed model to additional models of human-driving dynamics and other uncertainties and delays in real-world incident detection technologies. In addition, we seek to conduct a proof of concept to validate the model performance in real-world scenarios.


\bibliographystyle{IEEEtran}
\bibliography{neo}

\end{document}